\documentclass[12pt]{article}
\usepackage{graphicx}
\usepackage{epsfig}
\newcommand{\beq}{\begin{equation}}
\newcommand{\eeq}{\end{equation}}
\newcommand{\beqa}{\begin{eqnarray}}
\newcommand{\eeqa}{\end{eqnarray}}
\newcommand{\beqar}{\begin{eqnarray*}}
\newcommand{\eeqar}{\end{eqnarray*}}

\newcommand{\labell}[1]{\label{#1}}   %{\label{#1}\qquad_{#1}} %
\newcommand{\labels}[1]{\label{#1}} %{\vskip-2ex$_{#1}$\label{#1}}
\newcommand{\reef}[1]{(\ref{#1})}

\newcommand{\eg}{{\it e.g.,}\ }
\newcommand{\ie}{{\it i.e.,}\ }

\newcommand{\norm}[1]{\raise.3ex\hbox{:}#1\raise.3ex\hbox{:}}

\newcommand\prt{\partial}

\newcommand\tr{{\tilde r}}

\newcommand\ls{\ell_s}  % string scale
\newcommand\kt{{\rm K3}}
\newcommand\re{r_{\rm e}} %cvj has modified these a bit without rcm's permission!
\newcommand\rc{r_{\rm i}}
\newcommand\trc{\tilde{r}_{\rm e}}
\newcommand{\V}{{\rm V}}
\newcommand{\Vs}{{\rm V}_*}
\newcommand{\Vh}{{\rm V}_{\rm H}}
\renewcommand{\S}{{\rm S}}

\newcommand\gs{g_s} % string coupling
\newcommand\nl{N_1} %or {{\widetilde Q}_1}
\newcommand\nf{N_5} %or {Q_5}
\newcommand\dnl{\delta N_1}  % number of branes on shell
\newcommand\dnf{\delta N_5}
\newcommand\pnl{n_1}  % number of branes on composite probe
\newcommand\pnf{n_5}
\begin{document}

\setlength{\unitlength}{1mm}

\thispagestyle{empty}
\rightline{\small hep-th/0105159 \hfill McGill/01-04}
\rightline{\small\hfill DCPT-01/39}
\rightline{\small\hfill NSF-ITP-01-30}
\vspace*{2cm}

\begin{center}
%{\bf \LARGE Ode to the Enhan\c con}\\
{\bf \Large The Enhan\c con, Black Holes, and the Second Law}\\
\vspace*{1cm}

{\bf Clifford V. Johnson}$^{a,}$\footnote{E-mail: {\tt
    c.v.johnson@durham.ac.uk}} {\bf and Robert C.
  Myers}$^{b,}$\footnote{E-mail: {\tt rcm@hep.physics.mcgill.ca}}

\vspace*{0.2cm}

${}^a${\it Centre for Particle Theory}\\
{\it Department of Mathematical Sciences, University of Durham}\\
{\it Durham DH1 3LE, England}\\[.5em]

${}^b${\it Department of Physics, McGill University}\\
{\it Montr\'eal, Qu\'ebec, H3A 2T8, Canada}\\[.5em]

\vspace{2cm} {\bf Abstract}
\end{center}
We revisit the physics of five--dimensional black holes constructed
from D5-- and D1--branes and momentum modes in type IIB string theory
compactified on K3. Since these black holes incorporate D5--branes
wrapped on K3, an enhan\c con locus appears in the spacetime geometry.
With a ``small'' number of D1--branes, the entropy of a black hole is
maximised by including precisely half as many D5--branes as there are
D1--branes in the black hole.  Any attempts to introduce more
D5--branes, and so reduce the entropy, are thwarted by the appearance
of the enhan\c con locus above the horizon, which then prevents their
approach. The enhan\c con mechanism thereby acts to uphold the Second
Law of  Thermodynamics.  This result generalises: For each
type of bound state object which can be made of both types of brane,
we show that a new type of enhan\c con exists at successively smaller
radii in the geometry, again acting to prevent any reduction of the
entropy just when needed.  We briefly explore the appearance
of the enhan\c con locus in the black hole interior.

\vfill \setcounter{page}{0} \setcounter{footnote}{0}
\newpage

\section{Introduction}

The idea behind the microscopic evaluation of the entropy of a class
of charged black holes using D--branes is a simple one: First, one
constructs an arrangement of D--branes with the same asymptotic
macroscopic charges as the black hole. At weak coupling, this is a
relatively benign situation, and we can evaluate the degeneracy
associated to these macroscopic charges by counting the number of
microscopic arrangements of D--branes which give rise to the
configuration. Finally, we trust that since we have counted BPS
microstates at weak coupling, the journey to strong coupling will not
prevent us from associating the result with the entropy of
the black hole which is supposed to form in the limit.

This worked very well for the prototype case, presented in
ref.\cite{count}, of five--dimensional black holes in type~IIB string
theory. One wraps $Q_5$ D5--branes on a four--manifold ${\cal M}$,
which we shall take to lie in the $(x^6,x^7,x^8,x^9)$ directions,
leaving a string in the $x^5$ direction.  This string is combined with
$Q_1$ D1--branes, also lying in $x^5$.  Finally the $x^5$ direction is
compactified and $Q_P$ units of momentum are added along this circle
direction.  Thus the final configuration appears as a pointlike
object, in the five uncompactified directions $(x^0,\ldots,x^4)$,
which has three macroscopic charges $(Q_1,Q_5,Q_P)$ associated to it.

At strong coupling, (or large charges) there is a non trivial
back--reaction on the geometry and the resulting spacetime solution is
a five--dimensional extremal black hole, having a horizon of area
${\cal A}=8\pi G \sqrt{Q_1Q_5Q_P}$ (where $G$ is Newton's constant).
The microscopic count of the degeneracy of D--brane states giving rise
to the microscopic charges gives exactly the result $S={\cal A}/(4G)$
for the Bekenstein--Hawking entropy\cite{bekhawk}.

In the case where $\cal M$ is K3, there are complications which have
not been considered in the present black hole context.  It is
known\cite{bsv,anom} that wrapping a D$p$--brane on K3 induces a
negative unit of D$(p-4)$--brane charge in the unwrapped part of the
worldvolume. As this is a BPS object, this gives a corresponding
negative contribution to the tension of the wrapped
object\cite{sunil,tens}. At strong coupling (or large charges), where
the D--branes have a non--trivial back--reaction on the geometry, for
generic combinations of parameters it was shown in ref.\cite{enhance}
that this situation can give rise to regions of the naive spacetime
solution where the tension of the wrapped brane is unphysical. This is
simply because the volume of the K3 upon which the brane is wrapped
can vary as a function of position transverse to the branes. There are
positions where $\V_{\rm K3}$ drops below the stringy value of
$\Vs\equiv(2\pi\ls)^4$. Within such regions there are 
naked time--like singularities, resulting from the K3 volume shrinking
all the way to zero.

This is particularly problematic when branes which supposedly generate
the geometry have negative tension in the $\V_{\rm K3}< \Vs$  region.
The proposal of ref.\cite{enhance} was that the locus of points where
the tension drops to zero ({\it i.e.,} where $\V_{\rm K3}=\Vs$)
---called the ``enhan\c con''--- marks the end of the validity of the
naive spacetime geometry produced by those constituent branes, and the
region within must be excised and replaced by a different geometry
appropriate to the situation in hand: For the simplest case of just a
single constituent type of D--brane, the spacetime interior to the
enhan\c con is proposed\cite{enhance} to be {\it flat}.\footnote{For a
  series of studies of the consistency of this type of construction
  and how it fits with enhan\c con and brane physics, see ref.\cite{jmpr}.}
We shall see here that we
will need more interesting geometries in the interior, since we have
momentum, and both unwrapped and wrapped branes present, of which only
the latter care about the enhan\c con. In fact, as we shall see,
excision is not always necessary.

An issue which arises as an immediate consequence of considering the
K3 case of constructing these black holes is the fact that one can
choose parameters such that at strong coupling the enhan\c con appears
{\it outside} of the black hole horizon. Naively, this may
complicate some of the entropy counting story, and so
we carefully reconsider this case. In fact, while the microscopic
entropy story remains essentially unchanged, we will demonstrate
that the enhan\c con does play a crucial role in the physics of the
black holes. We find that its existence is {\it essential} to the
correct operation of second law of thermodynamics! This is a quite
satisfying additional facet of the connection, forged by D--branes,
between microscopic and macroscopic physics of black holes.

\section{The Consistency of Excision} 
\labels{match}

We start by considering the strongly coupled limit, where we have
non--trivial spacetime geometry. Using the conventions adopted for
example in refs.\cite{thesis,primer} the Einstein frame metric is:
\beqa ds^2=f_1^{-3/4}f_5^{-1/4}\left(-dt^2+dz^2+k\,(dt-dz)^2\right)
+f_1^{1/4}f_5^{-1/4}ds^2_{\kt}+f_1^{1/4}f_5^{3/4}\left(dr^2 +r^2\,
d\Omega_3^2 \right) \labell{metric}
\eeqa 
where $ds^2_{\kt}$ is the
metric on a K3 manifold with a fixed volume V, 
and 
\begin{equation}
d\Omega^2_3=d\theta^2+\sin^2\theta(d\phi^2+\sin^2\phi\,d\chi^2)
\end{equation}
is the metric on a round three sphere: $(t,r,\theta,\phi,\chi)$
constitute polar coordinates in the directions $(x^0,x^1,x^2,x^3,x^4)$,
and the K3 is in the $(x^6,x^7,x^8,x^9)$ directions.  The $x^5\equiv
z$ direction is compact with period $2\pi R_z$.  The dilaton and
Ramond--Ramond (R--R) fields are given by: 
\beq e^{2\Phi}={f_1\over f_5}\ 
,\qquad F^{(3)}_{rtz}=\prt_r f_1^{-1}\ ,\qquad
F^{(3)}_{\theta\phi\chi}=2\,r^2_5\,\sin^2\theta\sin\phi\ .
\labell{fields} 
\eeq
The harmonic functions are given by
\beq
f_1=1+{r_1^2\over r^2}\ 
,\qquad f_5=1+{r_5^2\over r^2}\ ,\qquad 
k={r_P^2\over r^2}
\labell{harmonic}
\eeq
where the various scales are set by
\beq
r_5^2=\gs\ls^2\,Q_5\ ,\qquad r_1^2=\gs\ls^2\,{\V^*\over\V}\,Q_1\ ,\qquad
r_P^2=\gs^2\ls^2\,{\V^*\over\V}\,{\ls^2\over R_z^2}\,Q_P\ ,
\labell{numbers}
\eeq
where $\Vs=(2\pi\ls)^4$ is the magic duality volume of the K3.
Newton's constant is given by $16\pi G = (2\pi)^7 \gs^2\,\ls^8$.  The
apparent singularity at $r=0$ is only a coordinate singularity. It is
actually only an event horizon with vanishing surface gravity and area
${\cal A}_H=4\pi^3 V R_z\,r_1r_5r_p$, measured in the Einstein frame.
These properties translate into a vanishing Hawking temperature and a
Bekenstein--Hawking entropy of $S=2\pi\sqrt{Q_1\,Q_5\,Q_P}$.

Of course, the integers $Q_1,$ $Q_5$ and $Q_P$ measure the asymptotic
charges associated with the electric and magnetic R--R fluxes and the
internal $z$--momentum, respectively. We also introduce integers
$N_1$ and $N_5$ to denote the number of D1--branes and D5--branes,
respectively, in the system. Of course, we have $N_5=Q_5$. However,
as discussed above, wrapping the D5--branes on K3 induces a negative
D1--brane charge. Hence we have $N_1=Q_1+Q_5$ or alternatively
$Q_1=N_1-N_5$.

Now the volume of the K3 manifold (measured by the string frame
metric ${\widetilde G}_{\mu\nu}{=}e^{\Phi/2}G_{\mu\nu}$) is 
\beq \V(r) =
{f_1\over f_5} \V \ , \labell{vr} \eeq 
where V is the asymptotic volume of
the K3. Now at the horizon, 
\beq \Vh\equiv \V(r=0)={r_1^2\over r_5^2}\V ={Q_1\over Q_5}\Vs
={N_1-N_5\over N_5}\Vs\ , \eeq
and so if $r_1<r_5$, then $\Vh<\V$. So we see that as long as
$r_1<r_5$, that the volume of the K3 is shrinking as we move in from
asymptotic infinity.  Note that in the case of interest,
$r_1<r_5$, that the local string coupling, \ie $\gs
e^\Phi=\gs(f_1/f_5)^{1/2}$, is also decreasing.
When we reach $\V(r)=\Vs$ at some radius, new
 physics will come into play, and this is the ``enhan\c con''
locus of ref.\cite{enhance}. This radius may be computed easily to be:
\begin{equation}
\re^2 =\gs\ls^2 {\Vs\over (\V- \Vs)}(2\nf-\nl)\ ,
\qquad\left\{\matrix{>0&{\rm for}\quad 2\nf>\nl\cr
<0&{\rm for}\quad 2\nf<\nl}\right. \ ,
\labell{enhanconoutside}
\end{equation}
where $\re^2<0$ simply indicates that the K3 volume reaches $\Vs$ inside
the event horizon. Therefore
we see that we can have the enhan\c con appearing either above
or below the horizon, depending upon our choices of parameters. As we
shall see, this will lead to very interesting physics.

\subsection{Matching for $\re^2>0$} \label{positive}

Now when the K3 volume reaches $\Vs$, at the enhan\c con radius,
$\re$, the wrapped D5--branes will be unable to proceed
supersymmetrically into smaller radius\cite{enhance}, due to the fact
that their effective tensions are going through zero there.  They are
therefore forced to form an enhan\c con sphere at radius $\re$.  Note
however that there is nothing to prevent the D1--branes and momentum
modes from moving inside of $r=\re$: They are not wrapped on K3 and
therefore do not care that it is approaching a special radius
there\cite{mytalk}. (We will illustrate these statements fully with a
probe computation later in section \ref{probe}.)  Hence while
eqns.~(\ref{metric}--\ref{harmonic}) provide a good supergravity
solution for $r>\re$, it seems that we need a different solution to
describe the interior (although it will become apparent in the sequel
that this need not be the case).

Naively, one may think that the interior solution should carry no D5--brane
charge, \ie the magnetic component of the R--R three--form
should vanish. However, later we will show that the D5--branes can
enter the region with $V(r)<\Vs$ and the black hole, if they are
appropriately ``dressed''. Hence we will consider a more general
extension of the supergravity solution given above in
eqns.~(\ref{metric}--\ref{harmonic}).  We introduce a shell at some
arbitrary radius $r=\rc$, which carries a fraction of the D--branes,
\ie $\dnl$ D1--branes and $\dnf$ D5--branes are uniformly distributed
over the three--sphere at $r=\rc$.  Thus the black hole which remains
in the interior  contains $\nl'=\nl-\dnl$ D1--branes and
$\nf'=\nf-\dnf$ D5--branes. Hence this black hole is characterised by
the charges: $Q_1'=\nl'-\nf'$, $Q_5'=\nf'$ and $Q_P$. The interior
supergravity solution then takes essentially the same form as above
\beqa ds^2&=&h_1^{-3/4}h_5^{-1/4}\left(-dt^2+dz^2+k\,(dt-dz)^2\right)
+h_1^{1/4}h_5^{-1/4}\,ds^2_{\kt}
+h_1^{1/4}h_5^{3/4}\left(dr^2 +r^2\,d\Omega_3^2\right)
\labell{metrici}\\
e^{2\Phi}&=&{h_1 /h_5}\ ,\qquad F^{(3)}_{rtz}=\prt_r h_1^{-1}\ ,\qquad
F^{(3)}_{\theta\phi\chi}=2\,\tr_5^2\,\sin^2\theta\sin\phi\ .
\labell{fieldsi} \eeqa The two new harmonic functions introduced here
are \beq h_1=1+{r_1^2-\tr_1^2\over \rc^2}+{\tr_1^2\over r^2}\ ,\qquad
h_5=1+{r_5^2-\tr_5^2\over \rc^2} +{\tr_5^2\over r^2}
\labell{harmonici} \eeq while $k$ remains as in
eqn.~\reef{harmonic}, since we do not leave any momentum on the
shell branes.  The new D--brane scales are set by 
\beq
\tr_5^2=\gs\ls^2\,Q'_5\ ,\qquad \tr_1^2=\gs\ls^2\,{\V^*\over\V}\,Q'_1\ 
.  \labell{numberi} \eeq 
Recall that these charges are chosen so that
$Q_5-Q_5'=\dnf$ and $Q_1-Q_1'=\dnl-\dnf$. Now the K3 volume at the horizon is
given by
\beq \Vh={Q'_1\over Q'_5}\Vs={N'_1-N'_5\over N'_5}\Vs\ . 
\labell{involum}
\eeq
Note that this volume may be bigger than $\Vs$, and that in particular
if $\tr_1>\tr_5$, $\V(r)$ grows as we move to smaller radii inside the shell.

The normalisation of the constants is chosen in eqn.~\reef{harmonici}
to ensure that the metric is continuous at $r=\rc$.  There is,
however, a discontinuity in the extrinsic curvature which can be
interpreted in terms of a $\delta$--function source of stress--energy
at $r=\rc$, using the standard Israel junction conditions \cite{junc}
--- see also ref.\cite{mtw}. A more complete discussion of this analysis
in the context of enhan\c con physics is given in
ref.\cite{jmpr}, so here we only sketch the calculations. 
The extrinsic
curvature of the $r=\rc$ surface is
\beq
K^\pm_{AB} ={1\over2}\,n_\pm^C\prt_C G_{AB} =\mp{1\over2}\sigma\,\prt_r G_{AB}
\labell{extrin}
\eeq
where $n_\pm=\mp\sigma\,\prt_r$ is the outward
directed unit normal vector, with $\sigma=G_{rr}^{-1/2}$. Then
defining the discontinuity in the extrinsic curvature across the
gluing surface, $\gamma_{AB} =K^+_{AB}+K^-_{AB}$, the surface
stress--tensor becomes
\beq
S_{AB}={1\over 8\pi G}\left(\gamma_{AB}-G_{AB}\,\gamma^C{}_C\right)\ .
\labell{stress}
\eeq
Note that as with all energy calculations in
string theory, the above analysis is performed using the Einstein frame
metric. Now a rather lengthy calculation leads to the following result:
\beqa
S_{\mu\nu}&=&{\sigma\over 16\pi G}\left({f_1'\over f_1}+{f_5'\over
    f_5} -{h_1'\over h_1}-{h_5'\over h_5}\right) G_{\mu\nu}
\nonumber\\
S_{ab}&=&{\sigma\over 16\pi G}\left({f_5'\over f_5}-{h_5'\over h_5}
\right)G_{ab}
\nonumber\\
S_{ij}&=&0
\labell{stressa}
\eeqa
where $\mu,\nu$ denote
the $t$ and $z$ directions, $a,b$ denote the K3 directions, and $i,j$
denote the angular directions along the $\S^3$ at the incision.

A few comments are in order here: 
\begin{itemize}
\item The surface tension along the angular directions vanishes. This
had to result since we are describing a BPS configuration and there
should be no stresses required to support a shell of D1--
and D5--branes at any radius $r=\rc$.

\item The tension in the K3 directions only depends on the D5--brane
harmonic functions. This is, of course, the expected result since
only the D5--branes wrap these directions.

\item The momentum appears nowhere in $S_{AB}$, again because this is
a BPS configuration and none of the momentum is supported by the shell.

\item 
  Finally the surface stress--energy in the $t$ and $z$ directions is
  determined by a single ``tension'', \beq T_{\rm eff}={\sigma\over
    16\pi G}\left({h_1'\over h_1}+{h_5'\over h_5}-{f_1'\over f_1}
    -{f_5'\over f_5}\right)\ .  \labell{tensa} \eeq This tension
  should be that of the effective strings formed by the wrapped
  D5--branes and D1--branes in the shell.  Note the units are that of
  (length)$^{-9}$.  This is as it should be. This is still a
  five--brane energy/(length)$^5$ with a further average over the
  area, $A_3$, of the $S^3$ at $r=\rc$. Taking $\rc$ large, one finds
  exactly as expected,
\beq T_{\rm eff}= \dnf {\tau_5 \over A_3}
  \left(1-{\Vs\over\V}\right) + \dnl {\tau_1\over A_3 \V}
  \labell{check} 
\eeq 
where $\tau_p=((2\pi)^{p}\ls^{(p+1)}g_s)^{-1}$ is the standard tension
of a single D$p$--brane \cite{gojoe}.  Hence we see two contributions:
the first is that of the wrapped D5--branes and the second coming from
the D1--branes in the shell.  Similarly at large $\rc$, the tension in
the K3 directions becomes $T_{\rm K3-eff}= {\dnf \tau_5 / A_3}$,
without the extra correction from the wrapping on K3. Further note
that this simplified large $\rc$ calculation can be extended to any
radius, where one finds that the shell acts as a source for the
metric, dilaton and R--R fields precisely as expected from the probe
brane action, as illustrated in ref.\cite{jmpr,neils}.
\end{itemize}

Now recall the expressions for the harmonic functions in
eqns.~\reef{harmonic} and \reef{harmonici}. One finds that for general
$\rc$, up to a positive coefficient 
\beq T_{\rm eff}\propto
\rc^2\left(\dnf-{\Vs\over\V}(\dnf-\dnl)\right)- g_s\ls^2{\Vs\over\V}
\left((2\nf-\nl)\dnf-\nf\dnl\right)\ .  \labell{tensb} \eeq
Hence we
have a positive tension as long as 
\beq \rc^2>\trc^2\equiv\gs\ls^2\,\Vs
{(2\nf-\nl)\dnf-\nf\dnl\over (\V- \Vs)\dnf+\Vs\dnl}\ .
\labell{bound} \eeq 
and the tension vanishes at precisely
$\rc^2=\trc^2$. The expressions simplify somewhat if we set $\dnl=0$,
\ie if the shell contains no D1--branes. In this case,
\beq
\trc^2=\gs\ls^2 {\Vs\over (\V- \Vs)}(2\nf-\nl)=\re^2\ .
\labell{oldbound} 
\eeq 
which, as we have indicated, corresponds precisely to the usual enhan\c con
radius.  On the other hand, it is straightforward to see from
eqn.~\reef{bound} that for $\dnl>0$, $\trc^2<\re^2$ and hence the
volume of K3 when the shell tension vanishes is smaller than the
self--dual value: $\V(\trc)<\Vs$.

\subsection{Probing the Black Hole} \labels{probe}

In this section we shall ask D1-- and D5--brane probes about their
view of the geometry we have studied in the previous sections. Both of
these types of probe are natural in this situation, since they
preserve the same supersymmetries. However, we will consider a
slightly more general calculation involving a composite probe brane
consisting of $\pnf$ D5--branes and $\pnl$ D1--branes.
It is important for the physics of the following that this composite
probe is in the D5--branes' Higgs phase.  That is, this composite
probe is {\it not} simply a collection of individual D5--branes and
D1--branes moving together,
but rather the D1--branes have been absorbed as instanton
strings lying along the $z$--direction in the D5--brane world-volume.
We regard these instantons to be maximally smeared over the K3
directions and that we have chosen the orientation of the vevs of the
hypermultiplets arising from 1--5 strings such that the instantons are
of maximal rank in the $U(\pnf)$ gauge theory.
In this phase, the composite probe brane is then a true bound state,
\ie the fields describing the relative separation of the branes in the
Coulomb phase, are all massive.

The effective action for the composite brane probe regarded as
an effective string becomes
\beqa
S&=&- \int_\Sigma d^2\xi\,\, e^{-\Phi(r)} (\pnf\tau_5 \V(r) + (\pnl-\pnf)\tau_1)
(-\det{g_{ab}})^{1/2} 
\nonumber\\
&&\quad+ \pnf\tau_5 \int_{\Sigma \times\rm K3}\! C^{(6)} 
+ (\pnl-\pnf) \tau_1\int_\Sigma C^{(2)}\ . 
\labell{probeaction}
\eeqa
where $\Sigma$ is the unwrapped part of the brane's
world--volume, with coordinates $\xi^{0,1}$.
We note in the above action that the wrapping of the
D5--branes on the K3 introduces negative contributions to both the tension
and two--form R--R charge terms\cite{bsv,anom,sunil,tens,enhance}.
Above $g_{ab}$ is the pull--back of the string--frame spacetime metric:
\begin{equation}
g_{ab}=
e^{\Phi/2}G_{\mu\nu}{\partial X^\mu\over \partial\xi^a}{\partial X^\nu\over
  \partial\xi^b}\ .
\end{equation}
The background fields in which the probe moves are those of the black
hole solution given in eqn.~\reef{fields}. The corresponding R--R
potentials may be written as
\begin{equation}
C^{(6)}=f_5^{-1}dx^0\wedge dx^5\wedge \varepsilon_\kt\ ,\quad
C^{(2)}=f_1^{-1}dx^0\wedge dx^5\ ,
\end{equation}
where $\varepsilon_\kt$ denotes the volume four--form on the K3 space
with (fixed) volume V. Note that these R--R potentials do {\it not}
vanish asymptotically. However, this gauge choice is convenient
because it eliminates a constant contribution to the energy which
would otherwise appear in the following calculation. We should also
mention that we adopt the conventions convenient for working with
supergravity solutions,
as described in ref.\cite{dielec}, so that the coefficients of the
Wess--Zumino terms in eqn.~\reef{probeaction} are $\tau_p$ including a
factor of $1/\gs$.

We will now choose static gauge, aligning the coordinates of the
effective probe string with the $x^5$ direction and letting it move in
the directions transverse to K3 while freezing and smearing the
degrees of freedom on the K3:
\begin{eqnarray}
\xi^0&=&x^0\equiv t\nonumber\\
\xi^1&=&x^5\equiv z\nonumber\\
x^i&=&x^i(t,z)\ ,\quad i=1,2,3,4\ ,
\end{eqnarray}

After a brief computation, the result can be written as the effective
Lagrangian $\cal L$ for a string moving in the $(x^1,x^2,x^3,x^4)$
directions: 
\beqa {\cal L}
&=&{1\over 2}\left(\pnf\tau_5 \V
  f_1+ (\pnl-\pnf)\tau_1 f_5\right) \left[{\dot r}^2-{r'}^2+k({\dot
    r}-r')^2
+r^2\left({\dot
      \Omega_3}^2-{\Omega'_3}^2+k({\dot\Omega_3} -\Omega_3')^2
  \right)\right]\ , \nonumber\\\label{probefive} \eeqa
where dot and prime are used to denote
$\partial/\partial t$ and $\partial/\partial z$, respectively.
The notation for the angular contributions is such that the derivatives
of the angular positions are contracted with the standard metric
on the unit $S^3$, \eg
${\dot\Omega}^2_3={\dot\theta}^2+\sin^2\theta({\dot\phi}^2
+\sin^2\phi\,{\dot\chi}^2)$. Notice that there is no
non--trivial potential, since supersymmetry cancelled the mass against
the R--R charge.

The effective tension of the probe is given by the prefactor in
eqn.~\reef{probefive}.  We can already see that there is the
possibility that the tension will go negative when $\pnf>\pnl$.

Putting in the definitions of the harmonic functions given in
eqn.~\reef{harmonic}, we get that the tension remains positive as long as
\begin{eqnarray}
\left(\pnf\tau_5 \V
  f_1+ (\pnl-\pnf)\tau_1 f_5\right) &>& 0\nonumber \\
\mbox{\it i.e.,  }\quad
r^2&>&\gs\ls^2\,\Vs{(2\nf-\nl)\pnf-\nf\pnl\over
(\V- \Vs)\pnf+\Vs\pnl}\ ,
\labell{condition}
\end{eqnarray}
where the lower bound
is actually the same as $\trc^2$ in eqn.~\reef{bound} with
the substitutions: $\dnl\rightarrow\pnl$, $\dnf\rightarrow\pnf$. It is
satisfying that this fits perfectly with the consistency condition we
derived from the supergravity solution with the composite shell in the
previous section. 

Let us consider some special cases of this result.  If we remove all
of the D5--branes, the result for pure D1--brane probes is quite
simple, as setting $n_5$ to zero in the above result gives:
\beq {\cal L}_{\rm D1}={1\over 2} \pnl\tau_1 f_5 \left[{\dot
    r}^2-{r'}^2+k({\dot r}-r')^2
+r^2\left({\dot
      \Omega_3}^2-{\Omega'_3}^2+k({\dot\Omega_3} -\Omega_3')^2
  \right)\right]\ , \labell{probeone} 
\eeq
This is a natural result: The D1--brane is not wrapped on the K3 and
so its tension is positive everywhere. It simply floats past the
enhan\c con radius on its way to the origin without seeing anything
particularly interesting there\cite{mytalk}.

Note that the result \reef{probeone} is the same as would have
been obtained in the case of probing for a $T^4$ compactification,
assuming that we consider only motion in the directions
transverse to the torus. Similarly in the case that
$\pnl=\pnf$, we get:
\beq {\cal L}={1\over 2} \pnf\tau_5 f_1 \left[{\dot
    r}^2-{r'}^2+k({\dot r}-r')^2 +r^2\left({\dot
      \Omega_3}^2-{\Omega'_3}^2+k({\dot\Omega_3} -\Omega_3')^2
  \right)\right]\ , \labell{pureprobefive} \eeq 
which is the same as the result for pure D5--brane probes in the case
where they are wrapped on $T^4$. 
The cancellation of the induced tensions from K3 wrapping and
non--trivial instanton number in constructing the bound state probe
provided a simple result: the wrapped D5--branes, when appropriately dressed
with instantons, can indeed pass through the enhan\c con shell.

If we instead remove all of the D1--branes, we just get the familiar
result of ref.\cite{enhance} that the probe, made of pure D5--branes,
hangs up at the enhan\c con radius $\re$. Now we discover that
ref.\cite{enhance}'s result is just a special case of a more general
result: whenever there are more D5--branes than D1--branes making up
the probe (\ie $n_5>n_1$), there is a generalisation of the enhan\c
con radius, $\trc^2$, where the composite probe will become
tensionless and must stop. Notice that this happens in a
``substringy'' regime where $V_{\rm K3}<\Vs$.

\bigskip
\bigskip
\bigskip

\section{The  Physics of the Black Holes} 
\label{fizzx}

\subsection{Constructing a Black Hole}

The probe results of the previous section now can be seen to highlight
the relevance of the supergravity solution we studied in
section~\ref{match}, where we placed some of the D5--branes inside the
black hole along with the D1--branes, despite the fact that sometimes the
enhan\c con radius appears outside the horizon.  To orient the discussion,
we consider making the black hole by beginning with a ``large'' enhan\c con
shell containing $\nf$ D5--branes and (adiabatically)
bringing in $\nl$ D1--branes from infinity,
as well as $Q_P$ momentum modes. As the latter play no role relevant to
the enhan\c con, we will simply assume that they are carried in to the origin
along with the first few D1--branes.
As emphasised in the probe discussion, since
the D1--branes are not wrapped on the K3, their tension remains
positive everywhere. Hence the D1--branes can simply pass through the
enhan\c con shell on their way to the origin\cite{mytalk}.

With a small
number of D1--branes (\ie $\nl<2\nf$), the enhan\c con radius
\reef{enhanconoutside} remains well away from the origin. Hence
na\"\i vely, one might think that no black hole is formed, rather
the D5--branes must remain fixed in the enhan\c con shell. However,
let us examine the interior solution (\ref{metrici}--\ref{harmonici})
when we choose $r_{\rm i}=\re$ so that the K3 volume
\beq
\V(r)={h_1\over h_5}\V
\labell{innerun}
\eeq
starts at $\Vs$ at the incision radius.
If $N'_1=N_1$ and $N'_5=0$ (\ie all of the D1--branes at the origin
and all of the D5--branes at the enhan\c con radius), then it is easy
to show that $\V(r)$ grows as the radius decreases inside the shell.
Since $\V(r)>\Vs$ in this region, there is no obstacle to moving some
of the D5--branes from the enhan\c con shell to the origin and forming
a black hole. As D5--branes are moved to the origin, the growth of the
K3 volume is suppressed and stops when $N'_5=N_1/2$ as can be seen from
eqn.~\reef{involum}. At this point, the volume at the horizon and throughout
the interior region is a fixed constant, \ie $\V(r<\re)=\Vs$. 

While this solution (with $N'_1=N_1$, $N'_5=N_1/2$) may seem to be a
limiting configuration, the black hole can absorb more D5--branes using
the bound states considered in the previous section. For example, illustrated
in eqn.~\reef{pureprobefive}, a bound state with $n_1=n_5$ has no problem
moving in a region where $\V(r)<\Vs$. Hence when the D1--branes move
in from infinity, rather than making a passive transit through enhan\c con
shell, some of these D1--branes can bind to D5--branes as instanton
strings and the resulting D1/D5 bound states can move to the origin.
In this way, a black hole can
be constructed which contains $N'_5\le N_1$ D5--branes. 
Note that from eqn.~\reef{involum}, we see $\Vh<\Vs$ for $N_1/2<N'_5<N_1$.
That is, we are able to construct black holes surrounded by a region
where the K3 volume is less than $\Vs$. 

{}From the above discussion, we conclude there are several different regimes:
For $0<\nl<\nf$, the black hole can only absorb a fraction of the total
number of D5--branes (up to $\nf'=\nl$) and so the black hole is naturally
dressed by an enhan\c con shell. For $\nf<\nl<2\nf$, the black hole can
absorb all of the D5--branes but there is still a region
where $\V(r)<\Vs$ outside of the horizon. Finally for $\nl>2\nf$, the 
black hole can again absorb all of the D5--branes and since $\re^2<0$
the K3 volume reaches $\Vs$ inside the event horizon.

In either of the last two cases, the supergravity solution is given in
eqns.~(\ref{metric}--\ref{harmonic}), taken for $0<r<\infty$.
We have a black hole with horizon area set by the product $Q_1Q_5Q_P$.
In the first case ($0<\nl<\nf$), there is an enhan\c con shell and so
we must introduce the interior solution (\ref{metrici}--\ref{harmonici})
to describe the region $r<\re$. Then we have an interior black hole
with horizon area set by the product $Q'_1Q'_5Q_P$.

We should emphasise that all of the black holes as well as the
intermediate configurations involved in their construction are supersymmetric.
Hence we can choose, if we wish, to leave extra D1--branes and D5--branes
outside the horizon and not contributing to making the black hole. 
Further, in the regime $0<\nl<\nf$, we may choose to put the excess
D5--branes in an enhan\c con shell around the black hole, or we may
place them in some distant region thereby essentially removing them from the
problem. We leave it to the reader to verify that
the area of the black hole is still given by precisely 
${\cal A}=8\pi G \sqrt{Q'_1Q'_5Q_P}$ in either case.

>From the discussion here, we conclude that enhan\c con physics does play a
role in the black holes when the number of D1--branes is small. We can
sharpen our understanding of the precise nature of this role by
carefully examining the formulae for the horizon area (or black hole entropy).

\subsection{The Second Law of Thermodynamics}

\begin{figure}[ht]
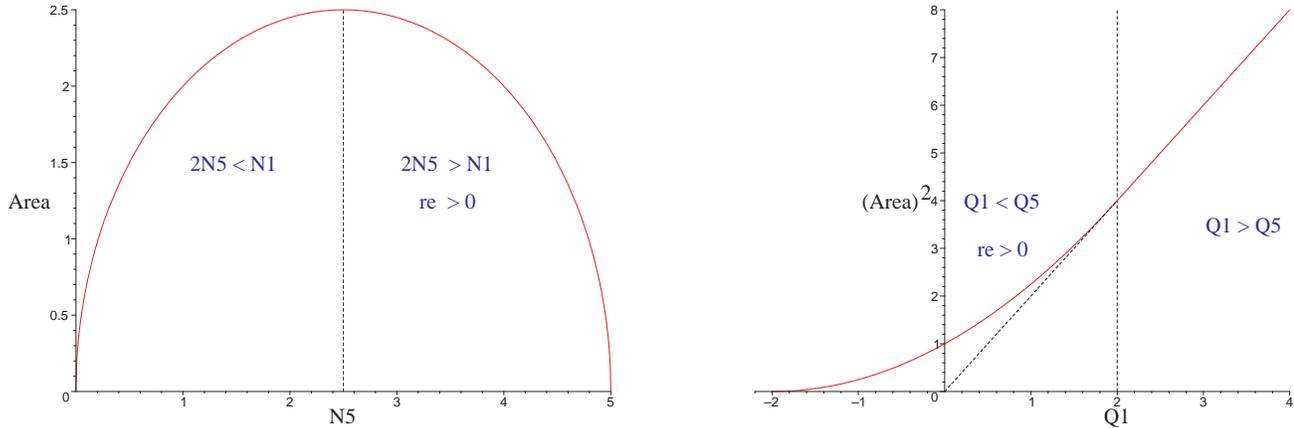

\centerline{\psfig{figure=areaplot1.epsi,height=2.5in}
\psfig{figure=areaplot2.epsi,height=2.5in}}
\caption{\small 
  On the left is the area as a function of $N_5$, for fixed $Q_P$ and
  $N_1$, which forms half of an ellipse.  As the number of
  five--branes increases past $N_1/2$, the area decreases. On the
  right is the square of the maximal horizon area as a function of
  $Q_1$, for fixed $Q_P\, ({=}1)$ and $Q_5\, ({=}2)$.  For $N_1>2N_5$,
  the (area)$^2$ increases linearly.  For $N_1<2N_5$, to maximise the
  area, one must use only $N_1/2$ of the available D5--branes (see
  left graph), and therefore the dependence on $N_1$ is quadratic.}
\label{area1}
\end{figure}

The entropy and area of the black holes which we construct are given by
the familiar formula
\beq
S= {{\cal A}\over 4G}= 2\pi \sqrt{Q_1Q_5Q_P}=2\pi \sqrt{(\nl-\nf)\nf Q_P}
\ .
\labell{sform}
\eeq
For fixed $\nl$ and $Q_P$, considering the dependence of the entropy
on the number of five--branes, we see that it gives a semi--ellipse,
as depicted in figure~\ref{area1}, on the left. Now while it is clear
that black holes form for any number of D5--branes, the {\it maximal
  entropy} black holes that we can make are those for which
$\nf=\nl/2$, or in other words $Q_1=Q_5$.  This is the apside of the
ellipse on the left in figure~\ref{area1}.

Hence if we wish to consider the maximum entropy that can be achieved
for a given set of parameters, $N_1$, $N_5$ and $Q_P$, we see that the
behaviour of this entropy changes at precisely $\nl=2\nf$.  The curve
on the right of figure~\ref{area1} shows the (square of the) maximal
entropy as a function of $Q_1$ for fixed $N_5$ and $Q_P$. For a
``large'' number of D1--branes ($\nl>2\nf$), the maximal area squared
is simply proportional to $Q_1$, as expected from eqn.~\reef{sform}.
However, for a ``small'' number of D1--branes ($\nl<2\nf$), the
entropy is maximised if only $\nf'=\nl/2$ of the available D5--branes
participate in the formation of the black hole. In this regime, we
have \beq {\cal A}_{\rm max}^2\propto \nl^2=(Q_1+Q_5)^2
\labell{rought} \eeq and so the curve becomes a parabola which only
reaches zero at $Q_1=-Q_5$.  Note that in this regime, the maximum
entropy is greater than one would calculate from eqn.~\reef{sform}.
Assuming the excess D5--branes have accumulated in an enhan\c con
shell around the black hole, the maximum entropy configuration
corresponds to precisely that where the K3 volume is frozen at $\Vs$
throughout the interior region.

Let us return to the curve on the left of figure~\ref{area1}.  Imagine
that we begin with a black hole with a ``large'' number of D1--branes.
It lies on the left hand side of the ellipse in the figure.  We may
now consider increasing the number of D5--branes in the system by
adding more one at a time. As we do so, the black hole moves up the
ellipse to the extremum at $\nf=\nl/2$. At this point, however, if we
were to add one more D5--brane, we we see that we will in fact {\it
  decrease} the horizon area, and hence the entropy of the resulting
system. We can in principle bring this D5--brane up to the black hole
horizon as slowly as we like. We seem, therefore, to have found a way
of reducing the entropy of the hole by an adiabatic process. This is a
violation of the second law of thermodynamics, which appears to be a
previously unconsidered flaw in the microphysics of black hole
thermodynamics, as represented by D--branes.

Happily, there is a very satisfying resolution of this problem. It is
precisely for this class of black holes that {\it the enhan\c con
  appears above the horizon}. So an attempt to bring our extra
D5--brane into the hole is thwarted by the fact that it will be forced
to stop at the enhan\c con radius $\re$ just above the
horizon\footnote{We are grateful to Joe Polchinski for a conversation
  in which this possibility arose.}.

We could bind the extra D5--brane with an extra D1--brane to bring it in,
but in this case $Q_1$ remains fixed while $Q_5$ increases. Thus dropping
in the D1/D5 bound state increases black hole entropy.

If we begin with a black hole on the right half of the ellipse
($\nl/2<\nf<\nl$), the enhan\c con again ensures that we cannot move further
to the right decreasing the horizon area by dropping D5--branes into
the black hole. These were configurations where the black hole is
already surrounded by a region where $\V(r)<\Vs$ and hence the extra
D5--branes are restrained from reaching the horizon by the enhan\c con
mechanism. 

However, we have seen in section \ref{probe} that D1/D5 bound states
can move through such regions where $\V(r)<\Vs$ and so we must still
investigate if we are able to decrease the entropy by sending in a
bound D1/D5 probe brane. Adopting the previous notation, let the probe
consist of a bound state with $\pnl$ D1--branes and $\pnf$ D5--branes.
Assuming that the black hole already contains many more of each type
of brane, \ie $\pnl,\pnf<<N_1,N_5$, dropping in such a probe would
cause an infinitesimal shift in the entropy (squared) given by
\beq
\delta S^2 =4\pi^2Q_P\left(\nf\pnl+(\nl-2\nf)\pnf\right)\ .
\labell{changes}
\eeq
Note that implicitly we are assuming $\nl,\nf,\pnl,\pnf>0$. Even so
the expression in parentheses has the potential to be negative
which would signal a decrease in the black hole entropy. However,
we found that this expression also appears
in the numerator of eqn.~\reef{condition} for the radius of vanishing
probe tension, but with the opposite sign!
Hence the probe--brane finds no obstacle to dropping inside the
horizon only in those situations where the entropy increases.
Precisely in those cases where second law would be violated, the
enhan\c con locus stands guard outside of the event horizon and the composite
branes are restrained from entering the black hole. Thus the enhan\c con
provides string theory with precisely the mechanism needed to
maintain consistency with the second law of black hole thermodynamics.

\subsection{Evaluating the Entropy in Gauge Theory}
 
Let us review briefly what the crucial elements of the entropy
counting argument are\footnote{See ref.\cite{adscft} for a review.}.
We assume that the scale of the K3 is much smaller than that of the
circle, so that we have an effective 1+1 dimensional gauge theory on
the effective D1--brane formed by wrapping the D5--branes and binding
it with D1--branes.  At strong coupling the theory will flow to a
conformal field theory in the infra--red. The important feature of the
conformal field theory is its central charge, which can be computed
from the gauge theory as proportional to $n_H-n_V$, the difference
between the numbers of hypermultiplets and the number of vector
multiplets.  Counting the bosonic parts, the D1--branes contribute
$\nl^2$ vectors and $\nl^2$ hypers, the latter coming from
$(x^6,x^7,x^8,x^9)$ fluctuations.  The D5--branes contribute $\nf^2$
vectors, but there are no massless modes coming from oscillator
excitations in the $(x^6,x^7,x^8,x^9)$ (K3) directions. There are, in
addition, 1--5 strings which give $\nl\,\nf$ hypermultiplets.
Evaluating the difference gives: $\nl\,\nf-\nf^2=Q_1Q_5$
hypermultiplets. Hence in total, there are $4Q_1Q_5$ bosonic
excitations and an equal number of fermions, since a hypermultiplet
contains four scalars and their superpartners.

In another language all that we have done is evaluated the dimension
the Higgs branch of the D5--brane moduli space of vacua, where the $\nl$
D1--branes can become instanton strings of the $U(N_5)$ gauge theory on the
world--volume of the D5--branes. The vacuum expectation values of the
1--5 strings is precisely what constitutes this branch. In this
language, the absence of hypers coming from the 5--5 sector
corresponds to the absence of Wilson lines on the K3 surface (since
the latter has trivial first homotopy class).  

Giving the D1--D5 bound string an overall momentum $P=Q_P/R_z$ in the
$x^5$ direction can be achieved in a number of ways, because of these
$4Q_1Q_5$ microstates and their fermionic superpartners, and in fact
the precise formula for this comes from the standard partition
function:
\begin{equation}
\left(\prod_{Q_P=0}^\infty{1+q^n\over 1-q^n}\right)^{4Q_1Q_5}=\sum_{Q_P=1}^\infty
\Omega(Q_P) q^{Q_P}\ ,
\end{equation}
where $\Omega(Q_P)$ is the degeneracy at level $Q_P$. (Recall that for
BPS excitations, the energy level (appearing in the partition
function) and the left (or right) moving momentum are equal.)  At
large $Q_P$ there is the result $\Omega(Q_P)\sim
\exp(2\pi\sqrt{Q_1Q_5Q_P})$.  So the entropy
$S\equiv\log(\Omega)=2\pi\sqrt{Q_1Q_5Q_P}$, which precisely matches the
strong coupling Bekenstein--Hawking result from the supergravity
solution listed below eqn.~\reef{numbers}.

Note that we have a mild paradox here. For $\nl<2\nf$, we know from
the analysis of the previous section that, at any given value of the
momentum, the entropy can be maximised by using only $\nl/2$ of the
D5--branes in the problem. So, on the one hand, it would seem that it
is favourable to Higgs the $U(\nf)$ gauge theory leaving massless
only a $U(\nl/2)$ subgroup. On the other hand,
the gauge theory cannot know this, since all of these supersymmetric
vacua are degenerate.
Therefore all black holes appear to be on the same footing from a field
theory point of view, despite
the fact that we can increase the entropy by not using all the
five--branes.

Clearly this puzzle is simply an artifact of
the thermodynamically peculiar situation that we are at zero
temperature while having a finite entropy. In such a situation, the
entropy strictly has a meaning as a degeneracy of ground states. Processes
which maximise the entropy require dynamics, and hence must take the system
(slightly) away from extremality. That is, we must slightly excite
the system in order that it can explore the phase space, and find
the configurations of maximal degeneracy or entropy as it settles back
to the ground state energy.
It would be interesting to study the effective couplings in the
full 1+1 dimensional model to see if they are consistent with the
system being able to ``seek'' the higher entropy configurations once
fluctuations are included.  Macroscopically, this must correspond to
the system being able to expel D5--branes in order to
increase its entropy, which is fascinating.

\section{Beyond the horizon} \label{inside}

Now we would like to consider the physics of the black hole interior,
the metric for which may be obtained by analytically continuing the
geometry of section \ref{match}.  In particular, one might wish to
consider the case where the volume of the K3 shrinks but does not
reach $\Vs$ until inside the event horizon, and so there is
enhan\c con physics inside the black hole.  

A simple choice of coordinates which cover the black hole interior\cite{gary}
is constructed as follows: As an intermediate step, change the radial
coordinate with $R^2=r_0^2+r^2$, which positions the horizon at 
$R=r_0$. However, we can consider interior points with $R<r_0$. We do
not present here the solution written in terms of $R$, rather we make
a second coordinate change $R^2=r_0^2-r^2$ --- note the sign ---
which puts the solution in a simple form. The interior metric becomes
\beqa ds^2=F_1^{-3/4}F_5^{-1/4}\left(dt^2-dz^2+k\,(dt-dz)^2\right)
+F_1^{1/4}F_5^{-1/4}ds^2_{\kt}+F_1^{1/4}F_5^{3/4}\left(dr^2 +r^2\,
d\Omega_3^2 \right)\ , \labell{metricinside}
\eeqa 
where now
\beq
F_1=-1+{r_1^2\over r^2}\ ,\qquad F_5=-1+{r_5^2\over r^2}\ ,
\labell{harmonicinside}
\eeq
where $k$ and the scales
$r_{1,5}$ are as before in eqns.~(\ref{harmonic},\ref{numbers}).  The
dilaton and R--R fields for the interior are given by:
\beq e^{2\Phi}={F_1\over F_5}\ ,\qquad F^{(3)}_{rtz}=-\prt_r F_1^{-1}\ 
,\qquad F^{(3)}_{\theta\phi\chi}=2\,r^2_5\,\sin^2\theta\sin\phi\ .
\labell{fieldsa}
\eeq

These coordinates are convenient, since they yield a structural form
for the interior solution similar to the one which we had outside of
the event horizon. Consequently, it is easy to compare results. Note,
however, that we have turned the radial coordinate ``inside out.''
The horizon is again positioned at $r=0$, and moving to larger values
of $r$ takes us further into the black hole interior. For example,
\beq
G_{\theta\theta}=(r_1^2-r^2)^{1/4}(r_5^2-r^2)^{3/4}
\labell{roundd}
\eeq
and so the three-sphere part of the geometry does indeed get smaller
as we move to larger values of $r$. The above metric element also suggests
that a curvature singularity may arise at $r=r_1$ or $r_5$. The regime
of interest here is $r_1<r_5$, in which case one does indeed encounter
a time--like singularity at $r=r_1$. We are also assuming that $r_P>r_1$ so
that there are no closed time--like curves in the interior geometry, as would
result if $G_{zz}\propto r^2_P/r^2-1$ became negative.

The volume of the K3 manifold (as measured by the string--frame
metric) is now given by
\beq
\V(r) = {F_1\over F_5} \V \ ,
\labell{vra}
\eeq
and we  see again that as long as $r_1<r_5$, 
the K3 volume continues to shrink as we move from the horizon
to larger values of $r$. Note that $\V(r)$ reaches zero at the singularity
$r=r_1$.

The enhan\c con radius in the new coordinates is given by the formula:
\begin{equation}
\re^2 =\gs\ls^2 {\Vs\over (\V- \Vs)}(\nl-2\nf)\ ,
\qquad\left\{\matrix{<0&{\rm for}\quad 2\nf>\nl\cr
>0&{\rm for}\quad 2\nf<\nl}\right.\ ,
\labell{enhanconinside}
\end{equation}
which is precisely the expected result compared to
eqn.~\reef{enhanconoutside}. That is, for $2\nf<\nl$ the enhan\c con
radius appears in the black hole interior.

For the case $2\nf<\nl$, we might expect that some of the wrapped
five--branes to hang at radius $\re$.  Hence consider an excision
construction where the above ``exterior'' solution is matched to
an ``interior'' solution at some radius $r=\rc$, as was done in section 2.
Note that with the present coordinates inside the event horizon the
new interior solution will be describing $r>\rc$.
As in section 2, the interior solution is characterised
by the charges: $Q_1'=\nl'-\nf'$, $Q_5'=\nf'$ and $Q_P$.
This solution then takes essentially the same form as above
\beqa ds^2&=&H_1^{-3/4}H_5^{-1/4}\left(dt^2-dz^2+k\,(dt-dz)^2\right)
+H_1^{1/4}H_5^{-1/4}\,ds^2_{\kt}
+H_1^{1/4}H_5^{3/4}\left(dr^2 +r^2\,d\Omega_3^2\right)
\nonumber\\
e^{2\Phi}&=&{H_1/H_5}\ ,\qquad F^{(3)}_{rtz}=-\prt_r H_1^{-1}\ ,\qquad
F^{(3)}_{\theta\phi\chi}=2\,\tr_5^2\,\sin^2\theta\sin\phi\ .
\labell{fieldsii}
\eeqa
The two new harmonic functions introduced here
are
\beq
H_1=-1+{r_1^2-\tr_1^2\over \rc^2}+{\tr_1^2\over r^2}\ ,\qquad
H_5=-1+{r_5^2-\tr_5^2\over \rc^2} +{\tr_5^2\over r^2}
\labell{harmonicii}
\eeq
while $k$ still remains as in eqn.~\reef{harmonic}, since we do not consider
the situation where the shell carries some of the momentum.
The new D--brane scales are set by 
\beq
\tr_5^2=\gs\ls^2\,Q'_5\ ,\qquad \tr_1^2=\gs\ls^2\,{\V^*\over\V}\,Q'_1\ 
.  \labell{numberii} \eeq 
Hence the R--R charges would indicate that the shell contains
$\dnf=Q_5-Q_5'$ D5--branes and $\dnl=Q_1+Q_5-Q_1'-Q_5'$ D1--branes.
Notice that even if the shell contains no D1--branes, 
the location of the singularity would be altered. The latter is now at
$r=\tr_1\rc/\sqrt{\rc^2+\tr_1^2-r_1^2}$.

Next, let us examine the stress tensor associated with the shell.
This is given by: \beqa S_{\mu\nu}&=&-{\sigma\over
  16\pi G}\left({F_1'\over F_1}+{F_5'\over F_5} -{H_1'\over
    H_1}-{H_5'\over H_5}\right) G_{\mu\nu}
\nonumber\\
S_{ab}&=&-{\sigma\over 16\pi G}\left({F_5'\over F_5}-{H_5'\over H_5}
\right)G_{ab}
\nonumber\\
S_{ij}&=&0
\labell{stressaa}
\eeqa
where as before: $\mu,\nu$ denote the $t$ and
$z$ directions; $a,b$ denote the K3 directions; and $i,j$ denote the
angular directions along the $\S^3$ at the incision. Note that with
our inside-out coordinates, the normal vectors switch their signs
compared to section 2. That is, $n_\pm=\pm\sigma\partial_r$ compared
to those introduced after eqn.~\reef{extrin}.

The important observation about this result, however, is that the
tensions appearing in this shell stress-energy are {\it negative}!
For example, one finds that the tension characteristic of the
K3 directions is now
\beq
T_{\rm K3-eff}\propto -\dnf\,{\tau_5/ A_3}\ .
\labell{negten}
\eeq
There is a similar minus sign in the effective string tension for the $t$
and $z$ directions, compared to eqn.~\reef{check}. However,
the effect of wrapping five--branes on the K3 space is still apparent,
so while this tension is negative near the horizon (\ie for small $\rc$),
it vanishes at the radius:
\beq \rc^2=\trc^2=\gs\ls^2\,\Vs
{(\nl-2\nf)\dnf+\nf\dnl\over (\V- \Vs)\dnf+\Vs\dnl}
\labell{boundinside} \eeq 
and becomes positive for smaller values of $r$ --- note that the
K3 tension remains negative in this region.
As a consequence, one cannot claim that the shell is constructed of
D5--branes and D1--branes (alone).  

These problematic results arise from the peculiar properties of the
black hole singularity\cite{don}, and are, in fact, 
entirely consistent with probe calculations\cite{dps}. The region
near the time--like singularity has a negative effective mass. Hence
any positive mass probe with the same charge as the singularity is
naturally pushed outward\cite{don}. That is, the trajectory of such a probe
heads towards the future Cauchy horizon at $r=0$. This effect can
be compensated for by also reversing the sign of the probe so that
the R--R forces are attractive while the effective gravitational force is
repulsive. Hence anti--branes become the natural probes
of the interior geometry, in the sense that the potential for
their motion vanishes\cite{dps}. This is consistent with the excision
construction above in that the characteristic tensions appearing
in eqn.~\reef{stressaa} would be positive if the parameters $\dnf$
and $\dnl$ were negative. That is, the stress--energy would be
well--behaved if the D5-- and D1--brane charges of
the solution inserted for $r>\rc$ were larger for the original exterior
solution, indicating that the shell was composed of anti--D5--branes
and anti--D1--branes. Note that the effective
string tension of the shell of anti--branes is positive for small values
of $\rc$ and vanishes precisely at the radius given in
eqn.~\reef{boundinside}. This latter result remains unchanged when
the signs of both $\dnf$ and $\dnl$ are flipped.

%{\it Let's get the hell out of here. It is just too weird inside the hole!
%                                                                       --cvj}

\section{Discussion} 
\labels{discuss}

An interesting and satisfying result of our investigation here is the
discovery that the enhan\c con radius is just the outer shell of an
onion--like structure arising when the supergravity solution is
constructed of multiple species of brane. There is a series of
concentric ``generalised enhan\c con'' shells where various
D$(p+4)$--D$p$ bound states become tensionless, for successively
smaller values of the K3 volume below the stringy value $\Vs$.  It is
immediately apparent that there are interesting cousins of these rich
structures to be found in the various U--dual situations involving
D--branes stretched between NS5--branes, fractional branes on a
collapsed two--cycle of K3 (with varying flux), and heterotic string
theory on $T^4$. For $p$ even  (type~IIA) it is clear that there is
enhanced gauge symmetry at these radii, while for $p$ odd 
(type~IIB), we get the enhanced two--form gauge symmetry associated to a
rich family of tensionless strings. These issues clearly deserve more
exploration.

The central issue of our investigation has been the interplay of the
physics of the enhan\c con with that of five--dimensional black holes.
After some initial thought, the enhan\c con mechanism seems to play no
essential role in the construction of the black holes. The only
situation where an enhan\c con must necessarily dress the black hole
exterior is in the regime $0<N_1<N_5$ (or $-Q_5<Q_1<0$), \ie when the
wrapped D5--branes make the largest contribution to the asymptotic
D1--charge.

Reconsidering the standard entropy formula \reef{sform}, we noted that
when the system contains a small number of D1--branes the entropy is
not maximised when all of the D5--branes are included in the black
hole. Rather, in the regime $0<N_1<2N_5$ (or $-Q_5<Q_1<Q_5$), the
entropy is maximised when the black hole only absorbs $N_5'=N_1/2$ of
the available wrapped D5--branes. This extended regime precisely
matches that where the enhan\c con locus \reef{enhanconoutside}
appears outside of the event horizon, and so it is natural to keep
some of the D5--branes at a distance away from the black hole.
Recall that for $N_5<N_1<2N_5$ (or $0<Q_1<Q_5$), in principle the
black hole could still absorb all of the D5--branes using D1/D5
bound states.

Na\" \i vely, this well--known entropy formula \reef{sform} allows us
to decrease the entropy in certain cases by throwing more wrapped
D5--branes into the black hole. Hence string theory might appear to allow
us to violate the second law of thermodynamics. Here, however, the
enhan\c con mechanism provides an elegant resolution of this paradox.
In precisely those configurations where the addition of D5--branes
would produce a violation of the second law, the enhan\c con locus
sits outside of the event horizon to prevent their infall. So we
discovered a fascinating interplay between the micro--physics of
branes and the macro--physics of supergravity black holes.

One is able to probe regions where the K3 volume is less than $\Vs$
with D1/D5 bound states in which D1--branes are absorbed as
instantonic strings on wrapped D5--branes. Thus there is a further
potential to violate the second law with these composite branes.
However, we found that the previous result generalised: There is a
family of generalised enhan\c con loci, inside the usual vanilla
enhan\c con, where these composite
branes become tensionless. They emerge in the
black hole exterior precisely in the regime where the infall of such
branes would reduce the horizon area.  Hence the black hole solutions
actually display a very delicate behaviour on macroscopic scales to
ward off the probes through the details of the micro--physics of
branes.

One may claim that supergravity shows a similar prescience of braney
physics in the matching calculations for the shells of section 2. That is,
the shell provides a source with precisely the correct R--R charges,
dilaton charge and stress--energy to match those of a shell of
D--branes, as was shown in great detail in ref.\cite{jmpr,neils}. We would
argue, however, that this result is essentially a result of
supersymmetry, \ie that supergravity knows that supersymmetric sources
with fixed R--R charges must have the characteristics we associate the
D--branes. The results of the matching calculations become much more
clouded for non--supersymmetric configurations\cite{jmpr}.

On the other hand, we must allow that the original entropy matching
calculations for these black holes\cite{count} are highly suggestive
that supergravity does have remarkable insight into at least certain
aspects of the brane physics, and of course the AdS/CFT
correspondence\cite{juan} confirms this. Specifically, how does
supergravity know to construct an event horizon whose area matches the
ground state degeneracy of the constituent branes? The enforcement of
the second law {\it via} the enhan\c con mechanism is a further
compelling demonstration that supergravity has a much deeper
``knowledge'' of the micro--physics of branes than meets the eye. Note
that this new phenomenon lies {\it outside} the AdS/CFT
correspondence, as the usual decoupling limit removes the enhan\c con.

Of course, the type IIB supergravity of interest here is embedded in a
superstring theory, and much of what we have found here relies on the
fact that supergravity retains a greater memory of the underlying
string theory than we might have had the right to expect: The enhan\c
con radius and its generalised cousins found here are supergravity's
mementos of the parent string theory. These radii are special, even in
supergravity, as we have seen here and as is emphasised also in
ref.\cite{jmpr}.  One might argue that supersymmetry may play an
essential role in producing our results, but the connection seems
somewhat obscure since the second law seems somewhat removed from
typical supersymmetry considerations.

Further insight might be gained by revisiting the supersymmetric
attractor equations which govern the exterior geometry of these black
holes\cite{attract}.  It would interesting to see if the attractor
flows ``know'' about the enhan\c con locus. Presumably the enhan\c con
must appear for the flows with $0<N_1<N_5$ (or $-Q_5<Q_1<0$) if they
are to avoid a repulson--like singularity.  Perhaps Denef's careful
analysis of ref.\cite{fred} can be extended to the present context to
show that supersymmetry naturally applies an excision at the enhan\c
con radius.

A related question would be to express our results in a U--duality
invariant way. That is, the relation $N_1=2N_5$ is distinguished in
the present analysis as a boundary between the regimes where the
maximal entropy black hole does or does not incorporate all of the
available D5--branes. Expressing this boundary using the U--duality
invariants\cite{invar} in the type IIB theory compactified on K3 seems
to be a nontrivial problem.

Just as the microscopic entropy counting can be extended to four
dimensional black holes\cite{fourdee}, one can also investigate the
role of the enhan\c con for these black holes. It seems that the
enhan\c con mechanism is again essential to enforcing the second law
for these black holes as well\cite{neils}. Of course, the structure of
the attractor flows and U--duality invariants would be even richer in
this setting.

As we commented at the end of section 3, supersymmetric configurations
provide a rather abstract framework for a discussion of
thermodynamics. Considerations such as maximising the entropy only
make sense in the context of dynamical processes which take the black
holes at least slightly away from extremality. Then one can imagine
that the system explores the full space of accessible states as it
settles back to being a supersymmetric black hole. Of course, the
non--extremal version of the black holes considered here is
well--known --- see, for example, refs.~\cite{juan,peet}.  In this
context, the application of the excision technique of section~2 may
produce ambiguous results\cite{jmpr}.  However, this is unlikely to be
an obstacle to studying the important question of how the enhan\c con
physics works to enforce the second law for these non--extremal black
holes.

It would be interesting to consider near--extremal black holes in the
regime $N_5<N_1<2N_5$ (or $0<Q_1<Q_5$).  Here, if the black hole
begins with all of the D5--branes at the horizon, its entropy is not
maximised. So we anticipate a new type of instability in this context,
leading to the expulsion of some of the
D5--charge from the black hole, possibly by the creation of
D5/anti--D5 pairs near the horizon. It would be interesting to see if
this is a classical instability or if it remains a quantum instability
similar to those arising in the discharge of Reissner--Nordstr\o m black
holes\cite{discharge}. It would be interesting to see if the local
stability analysis suggested in ref.\cite{local} might give some
insight into this question.  It would also be interesting to
understand this new expulsion mechanism in the microscopic framework
of the conformal field theory living on the D--branes.

Another interesting situation to reconsider for non--extremal black
holes is the case where the enhan\c con radius occurs inside the
horizon.  Inside of a non--extremal horizon, the enhan\c con locus will
be a time, not a place! Hence not only will an infalling D5--brane
{\it necessarily} reach the enhan\c con radius, it must pass through
this surface into the region where $\V(r)<\Vs$. At first sight, this
would seem a paradoxical situation, but we can propose a simple
resolution. As the D5--brane approaches the enhan\c con radius, it can
nucleate the creation of a D1/anti--D1 pair.  If the D1--brane binds
to the D5--brane, the bound state will have no problem in passing
beyond the enhan\c con radius, and the anti--D1--brane sees no
obstacle to passing this surface on its own. Note that as a result of
the time--like nature of the radius in this situation, the
non--extremal black hole interior is a effectively dynamical
background and ``energy'' conservation does not represent an obstacle
for such a pair creation process. Still it would be interesting to
study the infall quantitatively to confirm whether this
process and/or some other mechanisms come into play.

In the case of the interior of the extremal black holes considered
briefly in section 4, we uncovered essentially no new insights. As a
result of the peculiar structure of the spacetime geometry, the
physics of D--branes in the black hole exterior seems to be replaced
by the physics of anti--D--branes in the interior. It seems this even
extends to the appearance of an enhan\c con locus. However, one of our
initial goals had been to investigate whether there is a compelling
role for the enhan\c con in resolving time--like singularity appearing
inside the black hole.  Unfortunately, it seems that the answer to
this particular question is no. However, any possible disappointment
which we might have felt has been eclipsed by our excitement about the
myriad new avenues for interesting physics investigations uncovered
here. We hope that we have infected the reader with similar feelings.

\newpage

\section*{Acknowledgements}
Research by RCM was supported by NSERC of Canada and Fonds FCAR du
Qu\'ebec. That of CVJ was supported in part by the University of
Durham and the U.K. Particle Physics and Astronomy Research Council.
We would like to thank the Aspen Center for Physics for hospitality
during the initial stages of this project. CVJ would like to thank the
ITP, UCSB for hospitality during part of this work, and the organisers
of the ``M--Theory'' workshop for enabling him to participate.
Research at the ITP was supported in part by the U.S.  National
Science Foundation under Grant No.  PHY99--07949. We would like to
thank Neil Constable, Frederik Denef, Don Marolf, Greg Moore, Amanda
Peet, Simon Ross, Oyvind Tafjord and especially Joe Polchinski for
useful conversations.  We also thank Neil Constable for a careful
reading of the manuscript.

\end{document}